\documentclass{article} \begin{document}

\title{Quantum-electrodynamic model of the finite-size electron and
calculation of the fine-structure constant}

\author{E.P. Likhtman\\
All-Russian Institute for Scientific and Technical Information,\\
Russian Academy of Science, Moscow, Russia\\
e-mail: likhevg@zebra.ru}
\maketitle

\begin{abstract}

We propose a model of a relativistic string formed by a scalar complex
field, acting as electromagnetic field source. An axiosymmetric
solutions of the stationary equations for the scalar and
electromagnetic fields are found numerically. The mass $m$ is
calculated as a function of the charge $e$ and the magnetic moment
$\mu$ of the system. The resulting toroidal structure is interpreted
as an electron because the calculated ratio $e^3/(2mc^2\mu)$ coincides
with the fine-structure constant $\alpha=e^2/(\hbar\,c)\approx
e^3/(2m_ec^2\mu_e)$.

\end{abstract}


\section{Introduction}

Removing divergences in quantum electrodynamics by renormalization
procedure allows calculating multiloop contributions to electrodynamic
quantities \cite{QE}. As a result, the theory leads to an excellent
agreement with experimental data. For example, taking into account
radiation corrections into the electron's magnetic moment leads to an
agreement with experiment within 10 significant digits. However, the
fine-structure constant, which enters all calculations, is not
calculated within the theory. It is possible that one has to go beyond
electrodynamics in order to calculate it. The possible value
$1/({4\,\pi})$ of the unifield coupling constant in the grand
unification schemes at the energies from $10^{15}$ to $10^{19}$ GeV is
still under discussion (see for example \cite{RI}). In addition,
calculations connecting this constant with the fine-structure constant
in the energy interval of 17 orders of magnitude seem to be ambiguous
\cite{DG}. Alternative numerological approaches (see for example
\cite{BG}) give good agreement with experiment but fail to make any
predictions.

In quantum electrodynamics, the point-like electron is surrounded by a
cloud of electron-positron pairs and is actually a finite-size
structure. This fact suggests that it might be possible to use the
finite size structure as a ground state approximation in the
perturbation theory and thus avoid any divergences in the theory. The
"combined" models, where two or more material fields are connected by
the vector fields, are well known. For example, an electromagnetic
field connects electrons with nuclei in atoms, gluon fields connect
quarks in nucleons and mesons \cite{QU}. In contrast with this, in a
model presented here we have only one material field as a source of
electromagnetic field. The aim of the paper is to investigate the
soliton-like states of the system.

In the previous paper \cite{Li}, we proposed a classical model of the
finite-size axiosymmetric particle with charges and currents situated
on a disk. These charges and currents were assumed to be proportional
to the field potentials. Kaluza-Klein nonlinearity of electromagnetic
field \cite{KK1}, \cite{KK2} ensured the existence of stable solutions
and finite values of all observables in non-trivial stationary states.
However, the fine-structure constant calculated within the model
turned out to be of order one.

In this paper we propose a new model of extended electron in the
classical form of a flat ring current. Quantum fluctuations smear this
string out to a tore, while the charge $e$ and the magnetic moment
$\mu$ are invariant with respect to fluctuations. The basis of the
relativistic Lagrangian formalism is a scalar complex material field,
which acts as a source of the Maxwell electromagnetic field. The
interacting fields form a stable axiosymmetric soliton-like state. The
aim of the paper is validation of the proposed relativistic Lagrangian
formalism for the description of the electron as a finite-size
toroidal structure. The system of equations derived from this
formalism is solved numerically and the energy $E=m\,c^2$ of the stable state
is evaluated. The energy is a function of the model parameters $e$ and
$\mu$, but the dimensionless number $e^3/(2\,m\,c^2\,\mu)$ is independent
of them and must coincide with the fine-structure constant
$\alpha=e^2/(\hbar\,c)\approx e^3/(2m_ec^2\mu_e)$ (if one neglects the
electron's anomalous magnetic moment). The demonstrated agreement with
the experimental value serves as a validation of the proposed model.

\section {Lagrangian and parameterization of solutions}

We assume that the electromagnetic field is classical and the
Lagrangian has the form
\begin{equation} L_M=-\frac{1}{16\,\pi}\,\int
F_{\mu\nu}\,F^{\mu\nu}\,d^3x, \label{L_0}\end{equation} 
$$ F_{\mu\nu}=
\frac {\partial }{\partial x^{\mu} }\,A_{\nu } - \frac {\partial
}{\partial x^{\nu} }\,A_{\mu }. $$
The wave function $\psi$ is normalized as
\begin{equation}\frac{i\,e\,\lambda}{2}\int {
(\psi^*\frac {\partial }{\partial x_{0}}\psi-
 \psi\,\frac {\partial }{\partial x_{0}}\psi^*)\,d^3x}=e,
\label{l0}\end{equation}
where $\lambda$ is the fundamental length of the model determined later.
The interaction Lagrangian containing only linear terms in
potentials is 
\begin{equation} L_{int}=-\frac
{i\,e\,\lambda}{2}\,\int (\psi^*\frac {\partial }{\partial
x_{\nu}}\psi-\psi\,\frac {\partial }{\partial x_{\nu}}\psi^*)
\,A_{\nu}\,d^3x, \label{Li} \end{equation}
the Lagrangian of the "free" field $\psi$ has the form
\begin{equation}
L_0=\frac{e^2}{2\,\lambda}\,\int (
\lambda^2\frac{\partial }{\partial x^{\nu}}\psi^*\,
\frac{\partial }{\partial x_{\nu}}\psi\,-\psi^*\psi)\,d^3x.
\label{L0}\end{equation}

We note the parameters of the model are the charge $e$ and the
fundamental length $\lambda$. Note also that there are no bilinear in
electromagnetic potentials terms in Lagrangian components (1, 3, 4),
which are required by the gauge invariance. Thus, conservation of
electromagnetic current does not follow from Lagrange field equations.

We seek a solution for $\psi$, which is axiosymmetric (around
$z$-axis) and has a plane of symmetry $z=0$ in the cylindrical
coordinate system centered in the center of the charge. Conservation
of electromagnetic current can be ensured by choosing the wave
function to be an eigenfunction of the operator
$ie\lambda\,{\frac{d}{dx_0}}$ and operator $ie\lambda/r\,{\frac
{d}{d\phi}}$, such that the phase depends only on $t$ and $\phi$, and
the amplitude depends only on $r$ and $z$:
$\psi=1/(2\,\pi)^{1/2}\psi\left( r,z \right)\,e^{i\phi-i\,t/\lambda}$.

The magnetic moment $\mu=\mu_e$ can be obtained by setting
$\lambda=\mu_e/e=\hbar/(2\,m_e\,c)= 1.93\cdot 10^{-13}$ m, i.e. $4\,\pi$
times smaller than the Compton wave length of electron. The parameter
$\lambda$ can be interpreted as a ring radius of the string current at
which the current four-vector is light-like. Because the
fine-structure constant $e^3/(2\,m\,c^2\,\mu)$ we are calculating is
dimensionless, we set $e=1$, $\lambda=1$, $c=1$ and calculate the mass $m=E$.

Because $\psi^*\left( r,z \right)=\psi\left( r,z \right)$, the
currents components $J_r$ and $J_z$ are equal to zero, and the
electromagnetic field has only two non-zero components $A_0\left( r,z
\right)$ and $A_\phi\left( r,z \right)$. Taking all these assumptions
into account, calculating the integrals with respect to $\phi$ and
varying the Lagrangian over $A_0\left( r,z \right)$, $A_{\phi}\left(
r,z \right)$, and $\psi\left( r,z \right)$, we get the Maxwell equations
and the wave function equation:
\begin{equation} \nabla^2 A_0 \left(
r,z \right)= - 2\,\psi^2\left( r,z \right),\label{Mx0}\end{equation}
\begin{equation} \nabla^2 A_{\phi} \left( r,z \right)=
-2\,\psi^2\left( r,z \right)/r,\label{MxA}\end{equation} 
\begin{equation}
\left(-1/2\,\nabla^2+U\left( r,z \right)\right) \psi \left( r,z
\right)=0, \label{Shr}\end{equation} 
$$U\left( r,z \right) =\frac
{1}{2\,r^2}+ A_0\left( r,z \right)-\frac {A_{\phi}\left( r,z
\right)}{r}.$$
The amplitude of the wave function is normalized as
\begin{equation}\int \!{\psi^2\left( r,z \right)\,dS}=1,
\label{NS}\end{equation}
where $dS=r\,dr\,dz$.

Because of the symmetry, the boundary conditions at $z=0$ require that
all functions are even functions of $z$. The electric field potential
is an even function of $r$, and the wave function and the magnetic
potential are odd functions of $r$. At large $r$ and $z$, the wave
function should vanish, and the asymptotic behavior of $A_0$ and
$A_\phi$ is respectively determined by $e$ and $\mu$.

The wave function equation (\ref{Shr}) has the form of the stationary
Schr\"odinger equation with zero self-energy. The magnetic field
creates the potential well surrounded by the barrier created by the
electric field at large distances and by the "centrifugal" forces at
small distances. The depth of the well, which is necessary for
existence of a stable solution, can be achieved at a characteristic
radius of the current that is much smaller than the fundamental
length. This leads to a small value of the fine-structure constant.

The wave function decreases at large distances. Therefore, the
asymptotic behavior of the electromagnetic field potentials are the same
as those of the point-like particle. This in turn allows 
calculating the wave function asymptotic behavior:
\begin{equation} \psi =
\exp(-(8^{1/2}\,(r^2+z^2)^{1/4}))\,r^{1/4}/(r^2+z^2)^{1/2}\,(1+1.5468/
(r^2+z^2)^{1/4}+...). \end{equation}
The relatively slow decay of the wave function (in comparison with 
Gaussian decay $\exp (-x^2/2)$ for the harmonic oscillator wave function)
is due to the fast decay at the boundaries of
the potential well and its smallness in the vicinity of the barrier. This
fact validates the quasi-classical approximation for the ground state.
We give a quasi-classical estimate of the toroidal structure radius
$r_0$. The potential well $A_\phi/r=1/(2\,r_0^3)$ and kinetic energy
$(\pi/(2\,r_0))^2$ must be approximately equal. This implies $r_0=2/\pi^2$.
The energy of the bounded state is
\begin{equation} E=\int\partial L/\partial(\partial _0 \psi^*)\,\partial
_0\psi^*\,d^3x+\int\partial L/\partial(\partial _0 \psi)\,\partial
_0\psi\,d^3x-L_M-L_0-L_{int}\end{equation}
and taking the normalization condition into account, we obtain that
\begin{equation} E=1-L_M-L_0-L_{int}. \end{equation}
For $\psi$ and $A$ satisfying the equations of motion, we have 
$L_0+L_{int}=0$ and $L_{int}=-2*L_M$. This results in
\begin{equation}E=1+\frac{1}{2}\,\int \!{
\psi^{2}A_{\phi}/r\,d^3x}.\label{E0}\end{equation}
Integrating over $\phi$ leads to the final result for the
energy of the bounded state,
\begin{equation}E=1+\frac{1}{2}\,\int \!{
\psi^{2}\left( r,z\right)\, A_{\phi}\left(r,z\right)/r\,dS}.
\label{E}\end{equation}
So, in the quasi-classical approximation we find $E=\pi^6/16$ and
$1/\alpha=\pi^6/8\approx120$. The fact that such a crude estimate gives 
a very reasonable result is encouraging.

\section {Numerical solution algorithm}

We propose the following change of variables in  order  to  take
the   non-polynomial  (exponential)  form  of  the  wave  function
into account effectively:
\begin{equation} r=L_r\,f(\rho),\label{fr}\end{equation}
\begin{equation}z=L_z\,g(\zeta),\label{gz}\end{equation}
where $L_r$ and $L_z$ are the scale factors,  $f(\rho)$ and $g(\zeta)$
are some logarithmic functions  which  transform  infinite quadrant
$r=0..\infty$,  $z=0..\infty$  into  the  unit   square   $\rho=0..1$,
$\zeta=0..1$:
\begin{equation}
f(\rho)=s\,\rho+(-ln(1-\rho^t))^{(t-1)/t}/\rho^{t-2},\label{fr1}\end{equation}
\begin{equation}
g(\zeta)=(-ln(1-\zeta^{v}))^{(v-1)/v}/\zeta^{(v-2)}.\label{gz1}\end{equation}
After this transformation (at fixed s, t, and v) the wave function is
interpolated by polynomials, which are easy to integrate and
differentiate. The problem then split into two parts: 

1)calculating $L_z$ at fixed $L_r/L_z$, $f(\rho)$ and $g(\zeta)$, which
leads to a convergent solution and 

2) choosing an "optimal" solution by optimizing $L_r/L_z$, $f(\rho)$ 
and $g(\zeta)$.

\subsection {Calculation of the scale factor}
 
In order to calculate $L_z$, we note that the nonhomogeneous Maxwell
equations have nontrivial solution for any wave function in the right
hand side, whereas the homogeneous Schr\"odinger equation has only
trivial solution with the approximate potentials obtained from the
approximate solution of Maxwell equations. We construct an iterative
process by adding nonhomogeneous term proportional to $\psi_i$ to eq.
(\ref{Shr}) for $\psi_{i+1}$:

\begin{equation}
\left(-1/2\,\nabla^2 +U_i\left( r,z \right)\right)\psi_{i+1}\left( r,z
\right) +U_i\left( r,z \right)\psi_i \left( r,z\right)=0.
\label{Shr1}\end{equation}

We start with some initial value of the normalized wave function
$\psi=\psi_0$, solve nonhomogeneous Maxwell equations, and then find
the next iteration $\psi_{i+1}$ from eq. (\ref{Shr1}). The resonance
increase of the norm of $\psi_{i+1}$ (by 7-8 orders of magnitude)
indicates that the nonhomogeneous term is small in comparison with the
analogous term containing $\psi_{i+1}$, and therefore, the iterations
can be stopped (after each iteration, the solution is normalized
according to eq. (\ref{NS})). In order to calculate the eigenfunction
of the wave function equation, we vary the scale factor $L_z$ (at
fixed $L_r/L_z$) until we find the resonance in the norm of $\psi$.
The energy value is then calculated.

We put the described procedure to a test by calculating of the odd
wave function of the first exited state of the 1-d harmonic
oscillator. A few first odd moments of the probability density
coincided with the analytical values within a few tenths of percent
(due to the interpolation over 8 points).

\subsection {"Optimal" lattice}

The search of the resonance value of $L_z$ was repeated after changing
the ratio $L_r/L_z$ or the functions $f(\rho)$ and $g(\zeta)$. The
energy was found to depend (within a few percents) on the particular
choice of $L_r/L_z$, $f(\rho)$ and $g(\zeta)$, suggesting a need for
an increase of the number of parameters of the wave function. Such an
approach leads to a prohibitive calculation time. An alternative
approach used in this paper is to choose the ratio $L_r/L_z$ and
functions $f(\rho)$ and $g(\zeta)$ in such a way that the result
achieves its asymptotic value with a relatively small number of
parameters (of order of thousand) determining the wave function and
the potentials. For this purpose, we introduce additional requirements
on functions (\ref {fr},\ref {gz}), which reduce arbitrariness and
increase the accuracy of the result.

1) We require a balance of contributions to the norm of the wave
function from large and small distances along $r$ axis and a balance
of contributions from large distances between $r$ and $z$ axis. This
requires that the probability density as a function of $2\,\rho-1,
\zeta$ should be close to axiosymmetric function (i.e., look like the
up-side down cup). This is analogous to the requirement that a few
first moments should be zero (we limited ourselves with the dipole,
quadrupole and octopole moments):
\begin{equation}\int (2\,\rho-1)\,{
\psi^2\left( r,z \right)\,dS}=0, \label{DP}\end{equation}
\begin{equation}\int ((2\,\rho-1)^2-z^2)\,{
\psi^2\left( r,z \right)\,dS}=0, \label{QP}\end{equation}
\begin{equation}\int (2\,\rho-1)^3\,{
\psi^2\left( r,z \right)\,dS}=0. \label{OP}\end{equation}

2) We characterize the fields by their values on the lattice with
coordinates defined by the zeros of the Legendre polynomials in order
to use the Gaussian quadratures numerical integration method. We can
also estimate the accuracy of obtained solution by calculating the
probability of the total square deviation of the full solution from
the solution $\psi_4$ interpolated over each fourth point of the
lattice:
\begin{equation}\delta P_4=\int { (\psi\left( r,z
\right)-\psi_4\left( r,z \right))^2\,\psi^2\left( r,z \right)\,dS}. 
\label{5}\end{equation}
The minimum of this deviation, i.e. the good approximation of the solution
by  the  low order polynomial,  suggests a high precision of numerical
integration and differentiation.

3) We can impose one more condition on the ratio $L_r/L_z$: we assume
that the characteristic volume $L_r^2\,L_z$ is less sensitive to
numerical errors and require that its derivative with respect to the
ratio $L_r/L_z$ (for fixed $f(\rho)$, $g(\zeta)$ and resonance $Lz$)
is zero:

\begin{equation}
\frac {\partial }{\partial (L_r/L_z) }\,L_r^2(L_r/L_z)\,L_z(L_r/L_z)=0. \label{Lrz}\end{equation}

We should stress that conditions (\ref {DP}..\ref {Lrz}) are not
physical equations but serve as a criterion for the most accurate
solution at fixed lattice size. We impose these conditions one by one
and calculate the ratio $L_r/L_z$ and the energy $H$ for fixed
$f(\rho)$ and $g(\zeta)$. The stability of results (minimum of
relative dispersion of $L_r/L_z$) with respect to the choice of
conditions \ref{DP}..\ref {Lrz} is considered as an evidence of a good
choice for functions $f(\rho)$ and $g(\zeta)$.

The most time-consuming part of the calculation is solving the system
of linear equations for the field values on the lattice sites. The
calculation time grows as the cube of the number of lattice sites. It
also grows with the increasing precision of numerical differentiation.
Therefore we differentiate the wave function using its interpolation
over 9 or 11 points and electromagnetic potentials over 5 points
because they are smoother than the wave function. Numerical
integration using Gaussian quadratures is exact for polynomials of
order of twice number of points on the lattice. In this work we use
lattices $20\times 10$, $24\times 12$ and $28\times 14$ along the
respective $\rho$ and $\zeta$ axis.

\section {Numerical results}

We use transformations (\ref{fr}, \ref{gz}) and seek a resonance value
of the scale parameter $L_z$, which leads to a convergent solution of
Maxwell equations (\ref{Mx0}, \ref{MxA}) and the wave equation
(\ref{Shr}) for fixed functions (\ref{fr1},\ref{gz1}) and different
values of the ratio $L_r/L_z$. Using five additional criteria
(\ref{DP}..\ref{Lrz}) for the optimal lattice, we calculate five
different values of the ratio $L_r/L_z$. A small scatter of these
values $\delta L/L=\delta (L_r/L_z)/(L_r/L_z)$ is considered as a
criterion for the optimal choice of functions (\ref{fr1}, \ref{gz1}).
The table below lists the results for three lattices of different size
where the minimum value of $\delta L/L$ is below 3\%.

First two columns specify the number of lattice sites along $\rho$ and
$\zeta$ respectively and the number of points used for calculating
derivatives of function $\psi$. Third column contains mean square
deviation of smoothed solution from the exact one(\ref{5}). Next
column lists relative deviation of the ratio $L_r/L_z$ from its mean
value for 5 different optimal lattice criteria. Fifth column contains
main results of the paper: the product of twice the energy and
magnetic moment divided by the charge cubed.

\begin{center}
\begin{tabular}{|c|c|c|c|c|c|c|}
\hline

Lattice sites&Num. diff.&$\delta P_4$&$\delta L/L$&$2\,m\,c^2\,\mu/e^3$&$\bar{r}$&$\sqrt{\bar{z^2}}$\\
\hline
$20\times 10$ & 9 pt & 0.1576 & 2.1 \% &134.13$\pm$ 0.14 & 0.1362 & 0.0744\\
$24\times 12$ &11 pt & 0.0144 & 1.4 \% &136.55$\pm$ 0.20 & 0.1363 & 0.0735\\
$28\times 14$ &11 pt & 0.0073 & 0.3 \% &136.82$\pm$ 0.05 & 0.1363 & 0.0718\\
\hline\end{tabular}\end{center}
The experimental value of the quantity $2\,m_e\,c^2\,\mu_e/e^3$ is
$137.19$ (it is slightly higher than $1/{\alpha}=137.03$ because of
the anomalous magnetic moment of the electron). So, the calculations
on the $28\times14$ lattice agree with the experimental value within
few tenths of percent, and the discrepancy decreases with increasing
the lattice size. As expected, the characteristic size of the toroidal
structure is almost an order of magnitude smaller than the fundamental
length (see two last columns of the table, which list the ratios of
the average radius and height of the toroidal structure to the
fundamental length) and two orders of magnitude smaller than the
Compton length of electron. Its typical values are of order
$2\cdot 10^{-14}$ m. The correction to the second bounded state of the
toroidal structure in non-uniform proton field is 7 orders of
magnitude smaller than the second level energy of a hydrogen atom and
10 times smaller than the Lamb shift. Thus, the contribution from the
finite size of the totoidal structure is comparable with the radiation
corrections, at least at small energies.

\section{Conclusions}

We propose a relativistic Lagrangian formalism, which lead to a system
of equations for the complex scalar and electromagnetic fields.
Assuming axial and mirror symmetries and specific angular and time
dependence of the wave function, we obtain a system of three equations
for the stationary amplitude of the wave function and two
electromagnetic potentials as functions of two spatial variables. We
construct an iteration procedure for solving these equations on a
lattice and the algorithm for selecting an optimal lattice. The
numerical solution of this system allow calculating the energy (or the
mass) of the soliton-like state. We interpret the obtained toroidal
structure as an electron because the value of dimensionless constant
$e^3/(2\,m\,c^2\,\mu)$ coincides with the experimental value
$e^3/(2\,m_e\,c^2\,\mu_e)\approx e^2/(\hbar\,c)=\alpha$ within few
tenths of percent. Further increase of numerical precision is
desirable.

One of possible confirmations of the proposed Lagrangian formalism and
the analogous axial symmetry might be the construction of another
soliton-like solution with larger mass and smaller magnetic moment
that can be interpreted as $\mu$-meson.

The correspondence principle requires constructing the effective
Lagrangian of toroidal structure, investigating it's properties,
including gauge invariance, and solving the well known electrodynamic
problems on its basis. One could try to find electron stationary
states in external fields, in particular, to calculate bound states of
toroidal electron in the field of the proton (hydrogen atom).

I am grateful to I.V. Tyutin, B.L. Voronov and A.E. Likhtman for useful 
discussions.

\end{document}